\documentclass[runningheads]{llncs}
\usepackage{graphicx}
\usepackage[T1]{fontenc}
\usepackage{verbatimbox,lipsum}
\usepackage{subcaption}
\usepackage[utf8]{inputenc}
\usepackage{listings}
\usepackage{pifont}
\usepackage{float}
\usepackage{tablefootnote}
\newfloat{lstfloat}{htbp}{lop}
\floatname{lstfloat}{Listing}

\usepackage{amsmath}
\begin{document}
%


\title{Scholarly Knowledge Graph Construction \\from Published Software Packages}

%
%
\author{Muhammad Haris\inst{1}\orcidID{0000-0002-5071-1658} \and
Sören Auer\inst{2,1}\orcidID{0000-0002-0698-2864} \and
Markus Stocker\inst{2,1}\orcidID{0000-0001-5492-3212}
}

\authorrunning{Haris et al.}

\institute{L3S Research Center, Leibniz University Hannover 30167, Hannover, Germany\\
 \email {haris@l3s.de} \and
TIB---Leibniz Information Centre for Science and Technology, Germany
\email{\{markus.stocker,auer\}@tib.eu}}

\maketitle              

\begin{abstract}
The value of structured scholarly knowledge for research and society at large is well understood, but producing scholarly knowledge (i.e., knowledge traditionally published in articles) in structured form remains a challenge. We propose an approach for automatically extracting scholarly knowledge from published software packages by static analysis of their metadata and contents (scripts and data) and populating a scholarly knowledge graph with the extracted knowledge. Our approach is based on mining scientific software packages linked to article publications by extracting metadata and analyzing the Abstract Syntax Tree (AST) of the source code to obtain information about the used and produced data as well as operations performed on data. The resulting knowledge graph includes articles, software packages metadata, and computational techniques applied to input data utilized as materials in research work. The knowledge graph also includes the results reported as scholarly knowledge in articles. Our code is available on GitHub at the following link: https://github.com/mharis111/parse-software-scripts.
\keywords{Analyzing Software Packages \and Code Analysis \and  Abstract Syntax Tree \and Open Research Knowledge Graph \and Scholarly Communication \and Machine Actionability}
\end{abstract}
\section{Introduction}
\label{s:introduction}
Scholarly artefacts (articles, datasets, software, etc.) are proliferating rapidly in diverse data formats on numerous repositories~\cite{Hendler_2014}. The inadequate machine support in data processing motivates the need to extract essential scholarly knowledge published via these artefacts and represent extracted knowledge in structured form. This enables building databases that power advanced services for scholarly knowledge discovery and reuse. 

We propose an approach for populating a scholarly knowledge graph (specifically, the Open Research Knowledge Graph) with structured scholarly knowledge automatically extracted from software packages. The main purpose of the knowledge graph is to capture information about the materials and methods used in scholarly work described in research articles. Of particular interest is information about the operations performed on data, which we propose to extract by static code analysis using Abstract Syntax Tree (AST) representations of program code, as well as recomputing the scientific results mentioned in linked articles. We thus address the following research questions: 
\begin{enumerate}
    \item How can we reliably distinguish scholarly knowledge from other information?
    \item How can we reliably determine and describe the (computational) activities relevant to some research work as well as data input and output in activities?
\end{enumerate}
Our contribution is an approach---and its implementation in a production research infrastructure---for automated, structured scholarly knowledge extraction from software packages.

\section{Related Work}
\label{s:related-work}
Several approaches have been suggested to retrieve meta(data) from software repositories. Mao et al.~\cite{mao} proposed the Software Metadata Extraction Framework (SOMEF) which utilizes natural language processing techniques to extract metadata information from software packages. The framework extracts repository name, software description, citations, reference URLs, etc. from README files and represent the metadata in structured format. SOMEF was later extended to extract additional metadata and auxiliary files (e.g., Notebooks, Dockerfiles) from software packages~\cite{kelly}. Moreover, the extended work also supports creating a knowledge graph of parsed metadata. Abdelaziz et al.~\cite{Abdelaziz2020ADO} proposed CodeBreaker, a knowledge graph which contains information about more than a million Python scripts published on GitHub. The knowledge graph was integrated in an IDE to recommend code functions while writing software. 

A number of machine learning-based approaches for searching~\cite{husain2020codesearchnet} and summarizing~\cite{ahmad-etal-2020-transformer,iyer-etal-2016-summarizing} software scripts have been proposed. The Pydriller~\cite{PyDriller} and GitPython frameworks were proposed to mine information from GitHub repositories, including source code, commits, etc. Similarly, ModelMine~\cite{modelmine} was presented to extract and analyze models from software repositories. The tool is useful in extracting models from several repositories, thus improves software development. Vagavolu et al.~\cite{Vagavolu} presented an approach that generates multiple representations (Code2vec~\cite{le2014distributed}, semantic graphs with Abstract Syntax Tree (AST) of source code to capture all the relevant information needed for software engineering tasks.
\begin{figure*}[t!]
  \centering
  \includegraphics[width=0.8\textwidth]{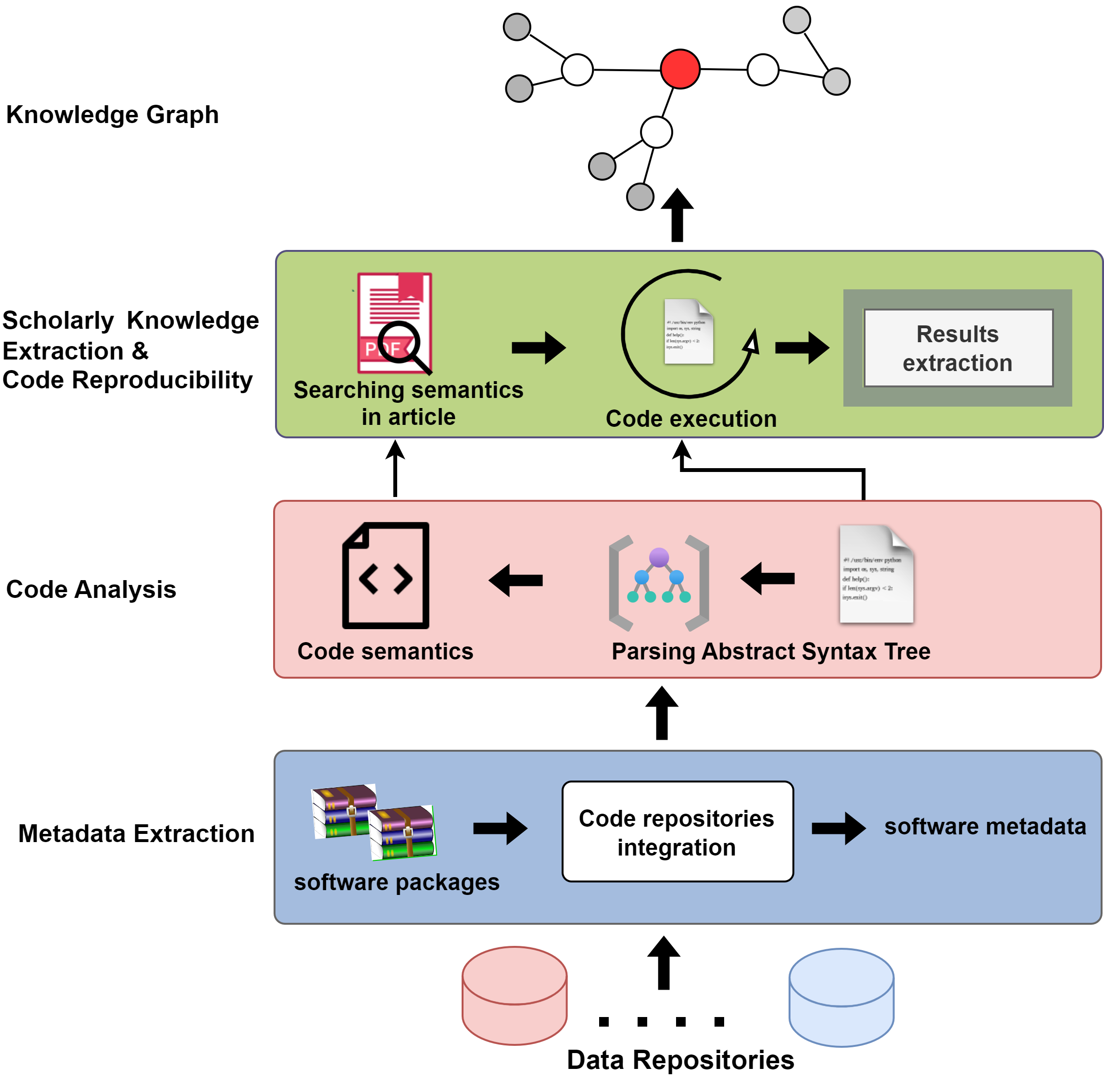}
  \caption{\small Pipeline for constructing a knowledge graph of scholarly knowledge extracted from software packages: 1) Mining software packages from data repositories using APIs; 2) Extracting software metadata by analyzing the APIs results; 3) Performing static code analysis using AST representations of software to extract code semantics; 4) Constrain code semantics to scholarly knowledge by matching the extracted information with article full text; 5) Recomputing the scientific results described in articles, by executing the scripts containing scholarly knowledge; 6) Knowledge graph construction with scholarly knowledge extracted from software packages.}
  \label{fig1}
\end{figure*}
\section{Methodology}
\label{s:methodology}
We now describe our proposed methodology for extracting scholarly knowledge from software packages and generating a knowledge graph from the extracted meta(data). Figure~\ref{fig1} provides an overview of the key components. We present the implementation for each of the steps of the methodology using a running example involving the article by Mancini et al.~\cite{mancini2021learning} and related published software package~\cite{mancini_flavia_2022_6997897}.

\subsection{Mining Software Packages}
We extract software packages from the Zenodo and figshare repositories by utilizing their REST APIs. The metadata of each software package is analyzed to retrieve its DOI and other associated information, specifically linked scholarly articles. Moreover, we use the Software Metadata Extraction Framework (SOMEF) to extract additional metadata from software packages, such as the software description, programming languages, and related references. Since not all software packages relate to scholarly articles explicitly in metadata, we also use SOMEF to parse the README files of software packages as an additional method to extract the DOI of linked scholarly articles.

\begin{figure*}[t!]
  \includegraphics[width=\textwidth]{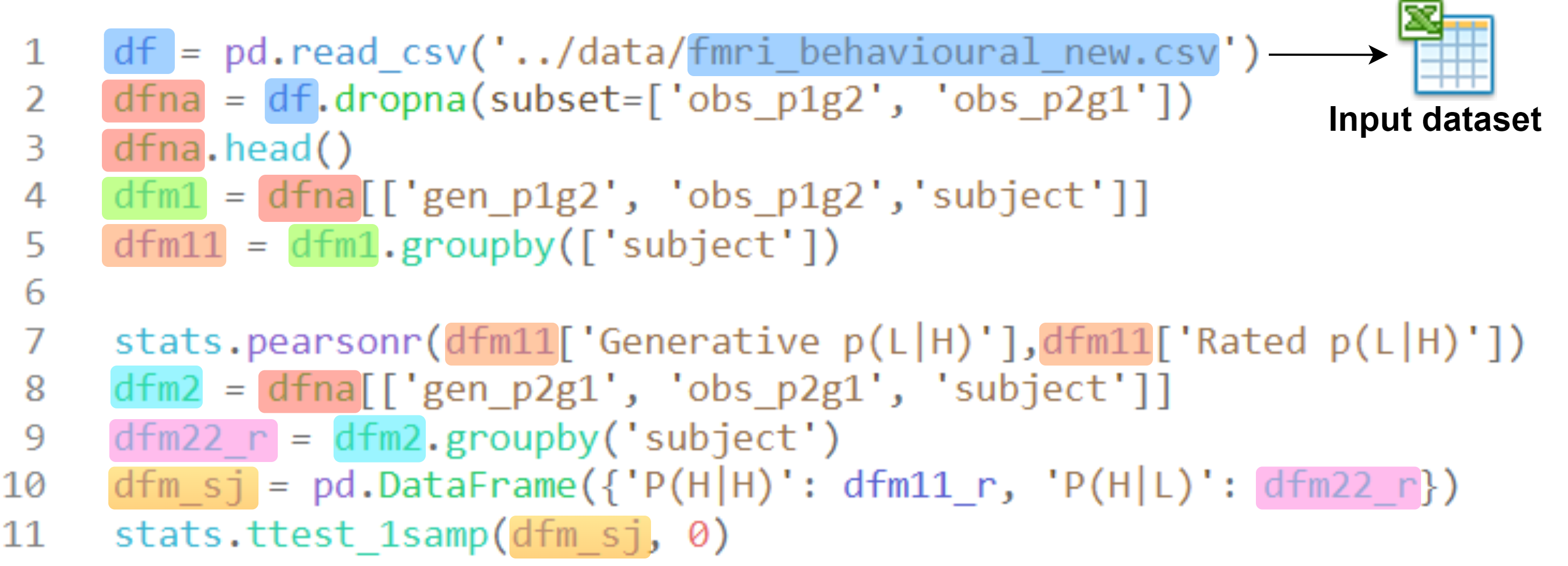}
  \caption{\small Static code analysis: Exemplary Python script (shortened) included in a software package. The script lines highlighted with same color show different procedural changes that a particular variable has undergone.}
  \label{fig2}
\end{figure*}

\subsection{Static Code Analysis}
\begin{figure*}[t!]
    \centering
  \includegraphics[width=\textwidth]{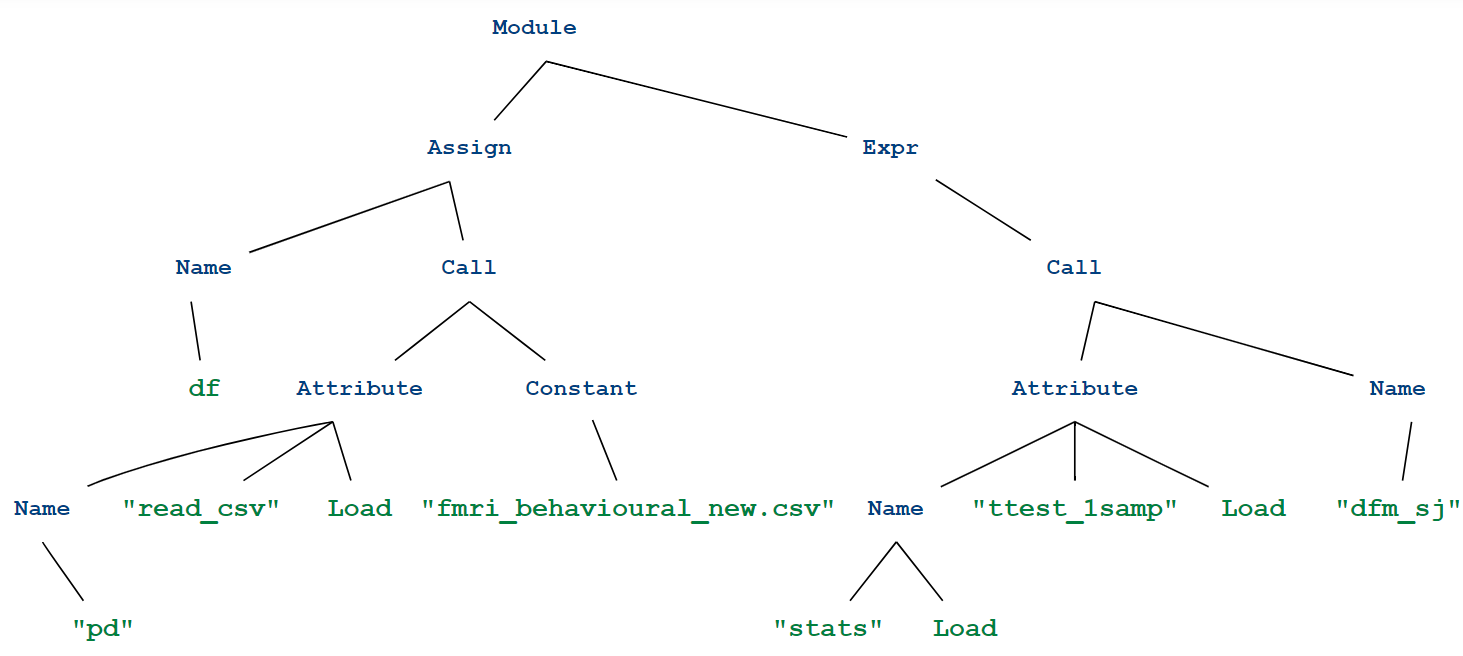}
  \caption{Abstract Syntax Tree (AST) of the script shown in Figure.~\ref{fig2}. For simplicity, the AST is shown only for lines 1 and 11. The child nodes of the Module node represent the operations that are performed in the respective lines of the script.}
  \label{fig3}
\end{figure*}

We utilize Abstract Syntax Tree (AST) representations for static analysis of Python scripts and Jupyter Notebooks included in software packages. Our developed Python-based module sequentially reads the scripts contained in software packages and generates the AST. The implemented methods and variables are represented as nodes in the tree, which facilitates the analysis of the code flow. Figure~\ref{fig2} shows the Python script included in the software package~\cite{mancini_flavia_2022_6997897}. The script illustrates an example in which the \texttt{fmri\_behavioural\_new.csv} data is loaded and two statistical hypothesis tests (\texttt{pearsonr} and \texttt{t-test}) are conducted on this data, respectively.
Figure~\ref{fig3} shows the AST of the Python script (Figure~\ref{fig2}) created using a suitable Python library. For simplicity, we show the AST of lines 1 and 11. We investigate the flow of variables that contain the input data, i.e., examining which operations used a particular variable as a parameter. With this analysis, we retrieve the series of operations performed on a particular variable. From our example, we conclude that \texttt{dropna}, \texttt{head}, \texttt{groupby}, \texttt{pearsonr} and \texttt{ttest\_ind} operations are executed on \texttt{fmri\_behavioural\_new.csv} data.

\subsection{Identifying Scholarly Knowledge}
\label{s:skg}
The information extracted via AST analysis of source code is not necessarily scholarly knowledge. In this work, scholarly knowledge is information expressed in scholarly articles. Hence, in this step we constrain the information obtained in the previous step to information mentioned in the articles linked to the analyzed software packages. We use the Unpaywall REST API\footnote{\url{https://api.unpaywall.org/v2/10.1101/2021.10.21.465270?email=unpaywall\_01@example.com}} to retrieve the document in PDF format. To identify the scholarly knowledge, we calculate the semantic similarity between code semantics and article sentences by employing a pre-trained BERT-based model to constrain words as scholarly knowledge that are semantically similar. First, we extract the text from PDF and remove stop words. Second, we arrange the sentences in bigrams and trigrams, because computing the similarity using a sentence-based corpus could lead to inefficient search. Next, we generate embeddings of keywords extracted from article text and software packages and use cosine-similarity to find words with similar meaning. From our example (Figure~\ref{fig2}), the extracted terms are searched in the article and we find that \texttt{fmri}, \texttt{pearsonr} and \texttt{t test} are mentioned in the linked article (Figure~\ref{fig4}(a, b)). Given the match, we assume that the extracted information is scholarly knowledge.

\subsection{Recompute the Research Results}
\begin{figure*}[t!]
  \centering
  \includegraphics[width=0.85\textwidth]{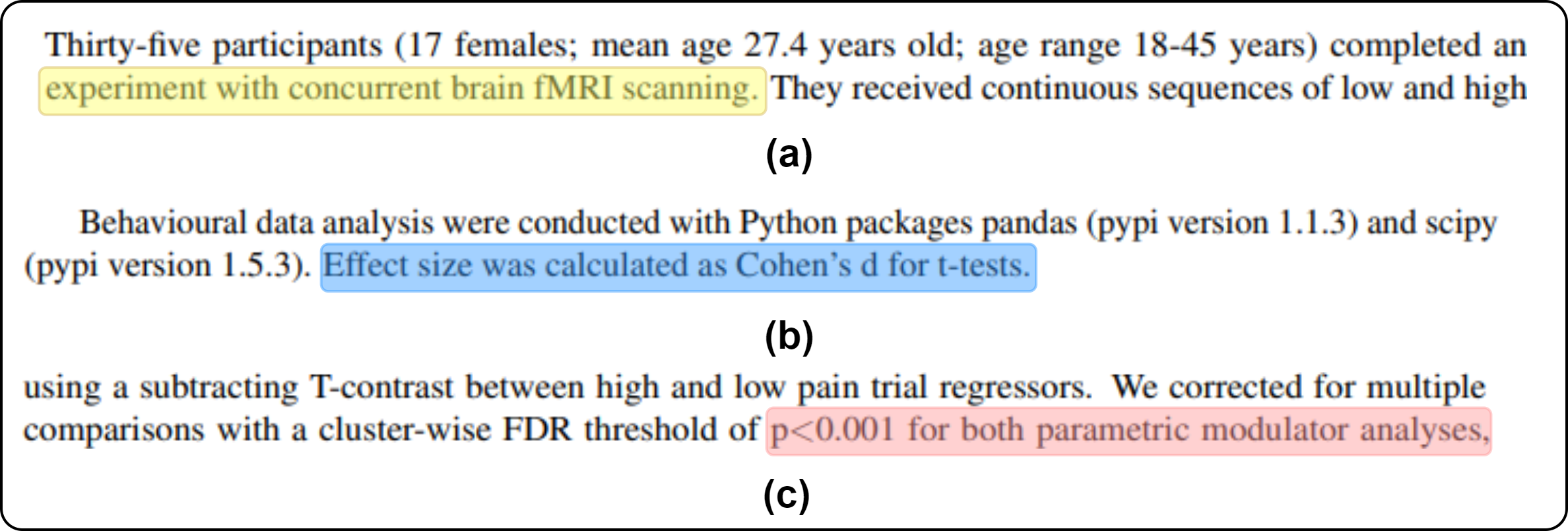}
  \caption{\small Snippets taken from the article (a) shows the name of the input dataset (b) \& (c) shows the statistical analysis (t-test) and the produced value $p < 0.001$, respectively.}
  \label{fig4}
\end{figure*}

For identified scholarly knowledge, we assume that the outputs of operations on data is information included in articles, and not necessarily stored as a separate dataset. In our running example, the output is a p-value and as we can see in Figure~\ref{fig2}, the output is merely printed to the console and presumably manually copied (and pasted into the article text). To enable rich descriptions of scholarly knowledge, we recompute the procedure outputs by executing the scripts that produce scholarly knowledge. For this, we develop a separate program to check the automatic execution of the code under consideration. If the code is not automatically executable then we introduce a human-in-the-loop step to execute the code. As different software packages may have been developed using varying versions of libraries, it is required to create a new virtual environment to execute scripts from the same package and prevent any conflicts between libraries. It is also necessary to check if the set of libraries required to execute the script are installed. If the libraries are not installed, we identify the required libraries using AST representations, and automatically install them in the virtual environment to execute the code under consideration. After successfully recomputing the code output, we assume that the computed outputs are correct and mentioned in the linked article. For our running example, we observed that the t-test returns a p-value (0.00088388), which is indeed mentioned in the paper (Figure~\ref{fig4}(c)).

\subsection{Knowledge Graph Construction}
\begin{figure*}[t!]
 \centering
  \includegraphics[width=\textwidth]{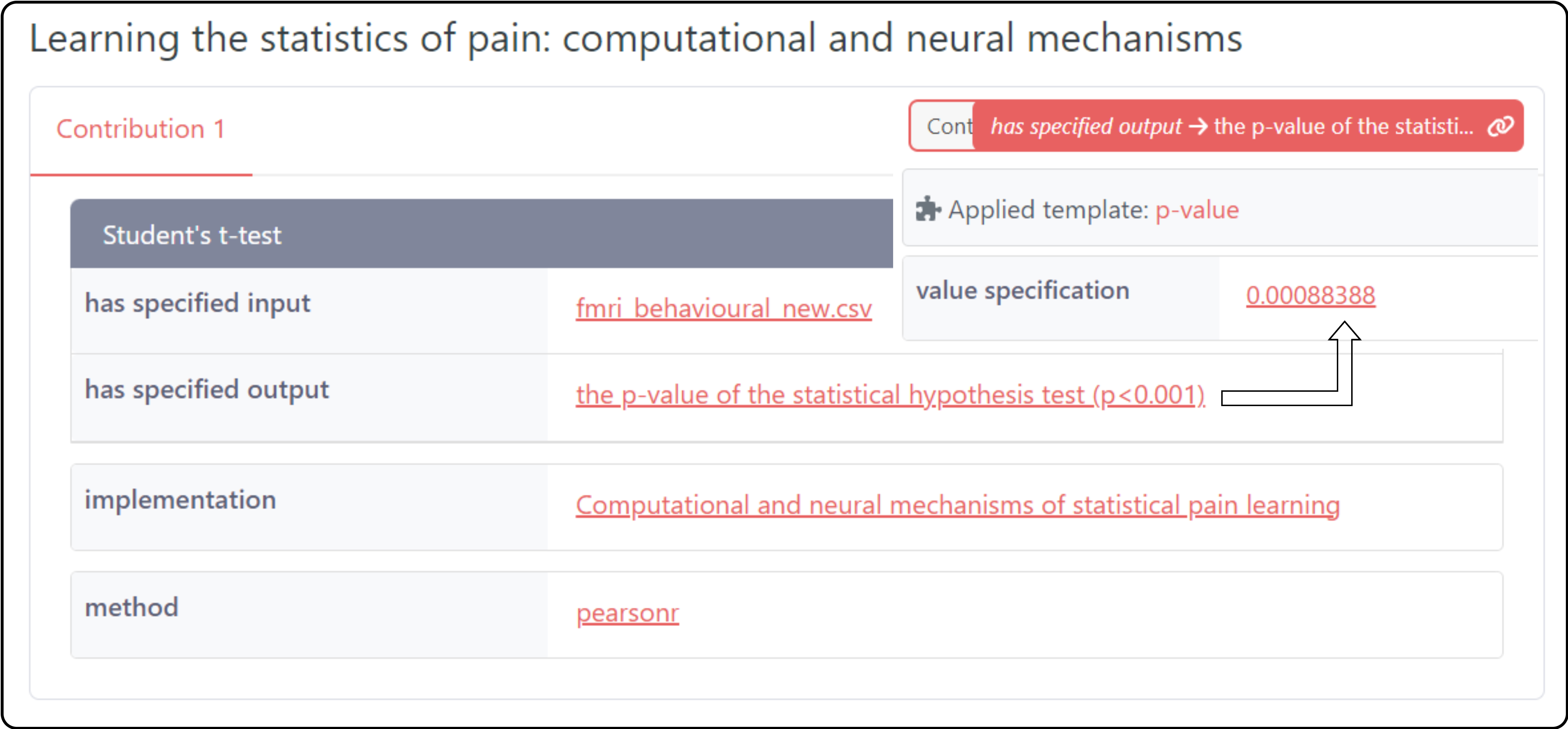}
  \caption{\small ORKG Paper showing the scholarly knowledge extracted from a software package, describing key aspects (e.g., statistical method used, its input and output data) of a research contribution of the work described in the article.} 
  \label{fig5}
\end{figure*}

Given the extracted scholarly knowledge, we now construct the knowledge graph or, in our case, populate an the Open Research Knowledge Graph (ORKG). First the meta(data) is converted into triples that are ingested into ORKG using its REST API. This conversion is guided by ORKG templates\footnote{https://orkg.org/templates}, which specify the structure of information types, their properties and value ranges. Hence, templates standardize ORKG content and ensure comparable semantics. To link the extracted scholarly knowledge with templates, we search for templates by operation name. The matching template is then utilized to produce ORKG-compliant data that can be ingested. For our running example (Figure~\ref{fig2}), we look for an ORKG template by searching ``t-test'' in the ORKG interface. We obtain a reference to the ORKG Student t-test template\footnote{\url{https://orkg.org/template/R12002}}, which specifies three properties, namely: has specified input, has specified output and has dependent variable. This template guides us in producing ORKG-compliant data for the scholarly knowledge extracted in the previous steps. Figure~\ref{fig5} shows the description of the paper in our running example in ORKG\footnote{\url{https://orkg.org/paper/R601243}, where readers can view the data also as a graph}. The metadata of the corresponding software package\footnote{\url{https://orkg.org/content-type/Software/R601252}} can also be viewed in ORKG.

\section{Validation}
Since we have employed AST to extract scholarly knowledge from software packages, it is essential to demonstrate that the proposed approach reliably produces the desired results. For this purpose, we compare the AST-based results with ground truth data extracted manually from software packages. To prepare the ground truth data, we have developed a Python script that iteratively reads software packages and scans each script to lookup all possible functions that load datasets (i.e., read\_csv, loadtxt, genfromtxt, read\_json, open). All scripts containing these functions are manually annotated for input data, operations performed on data, and output data, if any. To identify scholarly knowledge, the extracted information was searched in the full text of linked scholarly articles. Then, the manually extracted results were compared with the AST-based extracted results. The algorithmic performance is assessed using the Index of Agreement (IA)~\cite{Willmott1981}, a standardized measure to examine the potential agreement between results obtained using two different approaches. Its value varies between 0 and 1, indicating the potential error between the observed and predicted results. In total, we analyze 40 software packages and obtain an overall IA of 0.74. This result suggest an acceptable reliability of the proposed approach.

\section{Results and Discussion}
\label{s:discussion}
At the time of writing, more than 115,000 software packages are available on Zenodo and figshare, collectively. To expedite the execution, we consider packages of size up to 2 GB, i.e., 91,258 software packages. We analyze package metadata and the respective README files using SOMEF and find a total of 10,584 linked research articles. Our analysis focuses on Python-based software packages, of which there are 31,992. Among these, there are 5,239 Python-based implementations that are linked to 5,545 articles. We also observe that some software packages are linked to multiple articles. Table~\ref{table2} summarizes the statistics. To delve further into the structural and semantic aspects of these packages, we applied the AST-based approach and discovered that 8,618 software scripts (in 1,405 software packages) contain information about input datasets and methods executed on these datasets, and, if applicable, output datasets. A total of 2,049 papers are linked to these packages. As the the Open Research Knowledge Graph (ORKG) is designed purely for the publication of scholarly knowledge, we search the extracted knowledge in the full text of the linked articles (as explained in Section \ref{s:skg}) to identify scholarly knowledge. This step requires access to the full text of linked articles, and we found that out of 2,049 articles, 665 articles are closed access. Consequently, knowledge derived from packages linked to closed-access articles cannot be constrained as scholarly knowledge. Hence, we analyze the remaining 740 packages to obtain the scholarly knowledge. We describe the metadata of 91,258 software packages and 10,941 papers in ORKG. Out of total articles, 174 articles contain rich contributions i.e., information about datasets and operations performed on these datasets. The proposed AST-based approach is an alternative route to automatically produce structured scholarly knowledge. We suggest the extraction of scholarly knowledge from software packages, and our approach addresses the limitations of NLP-based approaches.

\begin{table*}[t!]
\caption{Statistics about the (scholarly) information extracted from software packages and added to ORKG as software descriptions, papers and their research contribution descriptions.}
\centering
\begin{tabular}{|l|l|}
\hline
  \textbf{Entity} & \textbf{Total} \\ \hline
\textit{Total software packages} & 115,000 \\ \hline
\textit{Software packages of size < 2 GB} & \multicolumn{1}{p{4cm}|}{91,258 (Zenodo: 87,423; figshare: 3,835)}    \\ \hline
\textit{Python-based software packages} & 31,992  \\ \hline
\textit{Software packages linked with articles} & \multicolumn{1}{p{5cm}|}{10,941, articles: 10,584}  \\ \hline
\textit{Python software packages, linked with articles} & \multicolumn{1}{p{5cm}|}{5,329, articles: 5,545}  \\ \hline
\textit{Packages containing information about datasets} & 1,405 \\ \hline
\textit{Packages containing scholarly knowledge} & 135, (articles: 174) \\ \hline
\end{tabular}
\label{table2}
\end{table*}

\section{Conclusions}
\label{s:conclusion}
We have presented a novel approach to structured scholarly knowledge production by extraction from published software packages. Based on the encouraging results, we suggest that the approach is an important contribution towards automated and scalable production of \emph{rich} structured scholarly knowledge accessible via a scholarly knowledge graph and efficiently reusable in services supporting data science. The richness is reflected by the fact that the resulting scholarly knowledge graph holds the links between articles, data and software at the granularity of individual computational activities as well as comprehensive descriptions of the computational methods and materials used and produced in research work presented in articles.

\section*{Acknowledgment}
This work was co-funded by the European Research Council for the project ScienceGRAPH (Grant agreement ID: 819536) and TIB--Leibniz Information Centre for Science and Technology.

\bibliographystyle{splncs04}
\bibliography{paper}
\end{document}